\documentclass[twocolumn,preprintnumbers,amsmath,amssymb]{revtex4}
\usepackage{graphicx}
\usepackage{dcolumn}
\usepackage{bm}
\usepackage{color}
\usepackage[final]{pdfpages}
\begin{document}
\title{A ferroelectric quantum phase transition inside the superconducting dome of Sr$_{1-x}$Ca$_{x}$TiO$_{3-\delta}$}
\author{Carl Willem Rischau$^{1}$, Xiao Lin$^{1,2}$, Christoph P. Grams$^{2}$, Dennis Finck$^{2}$, Steffen Harms$^{2}$,  Johannes Engelmayer $^{2}$, Thomas Lorenz$^{2}$,  Yann Gallais $^{3}$, Beno\^{\i}t Fauqu\'e$^{1,4}$, Joachim Hemberger$^{2}$ and Kamran Behnia$^{1}$}
\affiliation{(1) Laboratoire Physique et Etude de Mat\'{e}riaux (CNRS/ESPCI)\\ PSL Research University,  Paris, France\\
(2) II. Physikalisches Institut, Universit\"{a}t zu K\"{o}ln,  50937 K\"{o}ln, Germany\\
(3) Laboratoire Mat\'eriaux et Ph\'enom\`{e}nes Quantiques (CNRS-Universit\'e Paris Diderot) \\  75013 Paris, France\\
(4) IPCDF, Coll\`ege de France, 75005 Paris, France }
\date{January 9, 2017}
\begin{abstract}
 SrTiO$_{3}$, a quantum paraelectric\cite{Muller:1979}, becomes a metal with a superconducting instability after removal of an extremely small number of oxygen atoms\cite{Schooley:1964}. It turns into a ferroelectric upon substitution of a tiny fraction of strontium atoms with calcium\cite{Bednorz:1984}. The two orders  may be accidental neighbors or intimately connected, as in the picture of quantum critical ferroelectricity\cite{Rowley:2014}. Here, we show that in Sr$_{1-x}$Ca$_{x}$TiO$_{3-\delta}$ ($0.002<x<0.009$, $\delta<0.001$) the ferroelectric order coexists with dilute metallicity and its superconducting instability in a finite window of doping. At a critical carrier density, which scales with the Ca content, a quantum phase transition destroys the ferroelectric order.  We detect an upturn in the normal-state scattering and a significant modification of the superconducting dome in the vicinity of this quantum phase transition. The enhancement of the superconducting transition temperature with calcium substitution documents the role played by ferroelectric vicinity in the precocious emergence of superconductivity in this system, restricting possible theoretical scenarios for pairing.
\end{abstract}
\maketitle

\begin{figure*}
\includegraphics[width=0.8\textwidth]{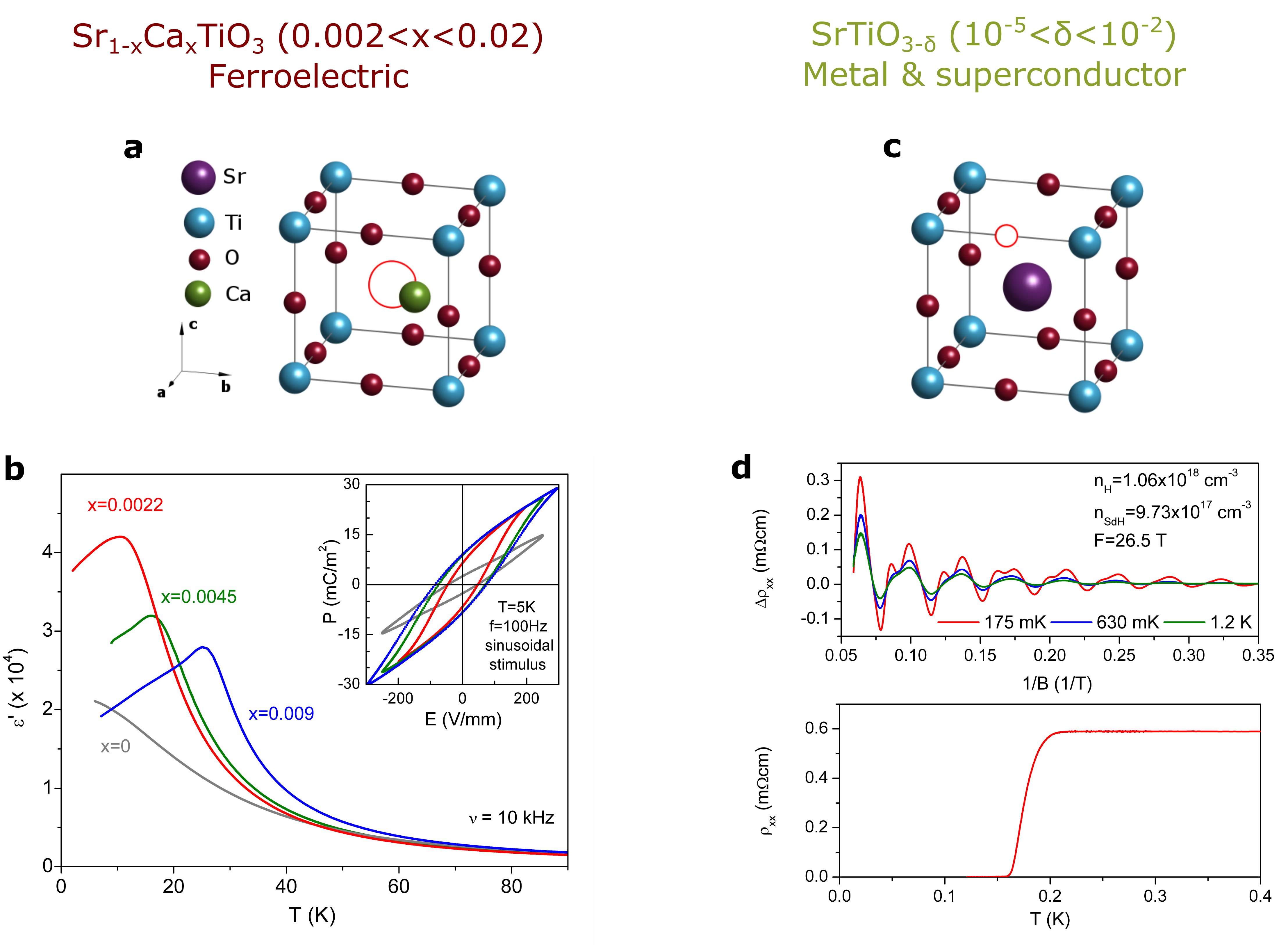}
\caption{\textbf{Emergence of ferroelectricity and metallicity by atomic substitution in SrTiO$_3$} \textbf{a)} Substituting strontium with isovalent and smaller calcium atoms leads to ferroelectricity. Smaller calcium atoms can take off-center positions and create local electric dipoles. Above a critical Ca threshold, a long-range ferroelectric order emerges below a Curie temperature.  \textbf{b)} Ferroelectricity in insulating Sr$_{1-x}$Ca$_{x}$TiO$_{3}$  documented by the temperature dependence of the real component of the dielectric permittivity, $\varepsilon$', for three different $x$. The maximum in $\varepsilon$' marks the Curie temperature. The inset shows $P(E)$ hysteresis loops at $T = 5$ K. \textbf{c)} Removing an oxygen atom introduces two n-type carriers. \textbf{d)}  Dilute metallic SrTiO$_3$ displays quantum oscillations of resistivity and a superconducting transition. The size of the Fermi surface according to the frequency of quantum oscillations matches the carrier density given by the Hall coefficient\cite{Lin:2013,Lin:2014}. }
\label{Fig1}
\end{figure*}

A perovskite of the $AB\textnormal{O}_3$ family, SrTiO$_3$ is a quantum paraelectric whose dielectric constant rises to $\sim 20000$ at low temperature\cite{Muller:1979}, but avoids long-range ferroelectric order. It becomes a metal by substituting Sr with La, Ti with Nb, or by removing O. It has been known for half a century that this metal is a superconductor at low temperatures\cite{Schooley:1964} and the superconducting dome has a maximum $T_c \sim0.4 $ K at $n\sim 10^{20}$ cm$^{-3}$\cite{Koonce:1967}. More recently, a sharp Fermi surface and  a superconductiong ground state have been found to persist down to a carrier concentration of 10$^{17}$cm$^{-3}$ in SrTiO$_{3-\delta}$\cite{Lin:2013,Lin:2014}. In this range of carrier concentration, exceptionally low compared to any other known superconductor, the Fermi temperature is an order of magnitude lower than the Debye temperature and the ability of phonons to form Cooper pairs is questionable\cite{Takada, Ruhman2016}.  Recent theories invoke fluctuations of the ferroelectric mode\cite{Edge:2015}, longitudinal optical phonons\cite{Gorkov2015} or plasmons\cite{Ruhman2016} as bosons responsible for generation of Cooper pairs.

In many unconventional superconductors, another electronic order competes with superconductivity. Here, the ferroelectric order is an obvious candidate\cite{Rowley:2014,Edge:2015}. It emerges in SrTiO$_3$ by isotopic substitution of $^{16}$O oxygen atoms with $^{18}$O \cite{Itoh:1999}, application of stress \cite{Uwe:1976} or substitution of a tiny fraction of Sr with Ca\cite{Bednorz:1984}. However, mobile electrons screen polarization and therefore only insulating solids are expected to host a ferroelectric order. Hitherto, as a paradigm, ferroelectric quantum criticality, in contrast to its magnetic counterpart, was deprived of an experimental phase diagram in which a superconducting phase and a ferroelectric order share a common boundary.

Here,  we produce such a phase diagram in the case of Sr$_{1-x}$Ca$_{x}$TiO$_{3-\delta}$. The main new observations are the following: i) Metallic Sr$_{1-x}$Ca$_{x}$TiO$_{3-\delta}$ hosts a phase transition structurally indistinguishable from the ferroelectric phase transition in insulating Sr$_{1-x}$Ca$_{x}$TiO$_3$; ii) The coexistence between this ferroelectric-like order and superconductivity ends beyond a threshold carrier concentration; iii) In the vicinity of this quantum phase transition, calcium substitution enhances the superconducting critical temperature and induces an upturn in the normal-state resistivity.

Figure \ref{Fig1} summarizes what we know about the emergence of ferroelectricity, metallicity and superconductivity in this system. When a small fraction of Sr atoms ($x>0.002$) is replaced with isovalent Ca,  Sr$_{1-x}$Ca$_{x}$TiO$_3$ becomes ferroelectric\cite{Bednorz:1984} with a Curie temperature steadily increasing with Ca content in the dilute limit $0.002<x<0.02$\cite{Bednorz:1984,Kleemann:1988,Bianchi:1994,Kleemann:1997,Kleemann:2000}. Macroscopic polarization below the Curie temperature has been observed in dielectric and linear birefringence measurements and found to build up in the plane perpendicular to the tetragonal axis along the $[110]$ and $[1\bar{1}0]$ directions\cite{Bednorz:1984,Kleemann:1997}. Fig.\ref{Fig1}b  presents the temperature dependence of the real part of the dielectric permittivity $\varepsilon'$ in our Sr$_{1-x}$Ca$_x$TiO$_3$ single crystals at three different Ca contents ($x$=0.0022;0.0045;0.009), obtained by measuring their complex conductivity. In the Ca-substituted system, there is a peak in the low temperature permittivity at the Curie temperature, $T_C$ and below $T_C$, the polarization $P$ shows a hysteresis loop.

In SrTiO$_3$, electric permittivity increases steadily with decreasing temperature and saturates at low temperatures due to quantum fluctuations attaining a magnitude as large as to 20000$\epsilon_{0}$\cite{Muller:1979,Hemberger:1996}. The ionic radius of Ca (0.99 {\AA})is smaller than Sr (1.12 {\AA}) and CaTiO$_3$ loses its cubic symmetry at 1600 K and suffers multiple structural transitions\cite{Carpenter:2006} without becoming ferroelectric. Its permittivity saturates to 350$\epsilon_{0}$\cite{Ang:2001}. The stabilization of the ferroelectric order in dilute Sr$_{1-x}$Ca$_{x}$TiO$_3$ is restricted to a narrow window in the dilute limit of Ca content($0.002<x<0.02$) within a highly polarizable matrix. There are two alternative ways to picture this order\cite{Kleemann:2000}. In the first one, it is driven by dipole-dipole interaction between off-center Calcium atoms, which form polarized clusters growing in size and percolating at the Curie temperature. In the second picture, the ferroelectric order is stabilized because the quantum fluctuations of the host matrix are pinned by Ca sites. A transverse Ising model with appropriate parameters\cite{Wang:1998} can reproduce the  critical doping for the emergence of ferroelectricity as well as the subsequent increase in the Curie temperature with $x$\cite{Bednorz:1984}.

By heating SrTiO$_{3}$ in vacuum, oxygen atoms are removed and mobile carriers are introduced\cite{Spinelli}. A metal-insulator transition\cite{Mott1990} is expected above a threshold carrier concentration\cite{Edwards}, which is exceptionally low because of the long Bohr radius\cite{Lin:2013}.  At a carrier density of 10$^{17}$cm$^{-3}$, several orders of magnitude above the expected threshold of metal-insulator transition, there is a single Fermi sea\cite{Behnia} (and not a collection of metallic puddles). This picture is based on the observation of quantum oscillations (Fig.\ref{Fig1}d), with frequencies\cite{Lin:2013,Lin:2014,Allen2013} matching the carrier density expected from the magnitude of Hall coefficient. This dilute metal is subject to a superconducting instability\cite{Schooley:1964,Koonce:1967,Lin:2013,Lin:2014}. The aim of this study is to find what happens to  metallicity and ferroelectricity when one removes oxygen and substitutes Sr with Ca\cite{DeLima:2015}.

\begin{figure*}
\includegraphics[width=0.9\textwidth]{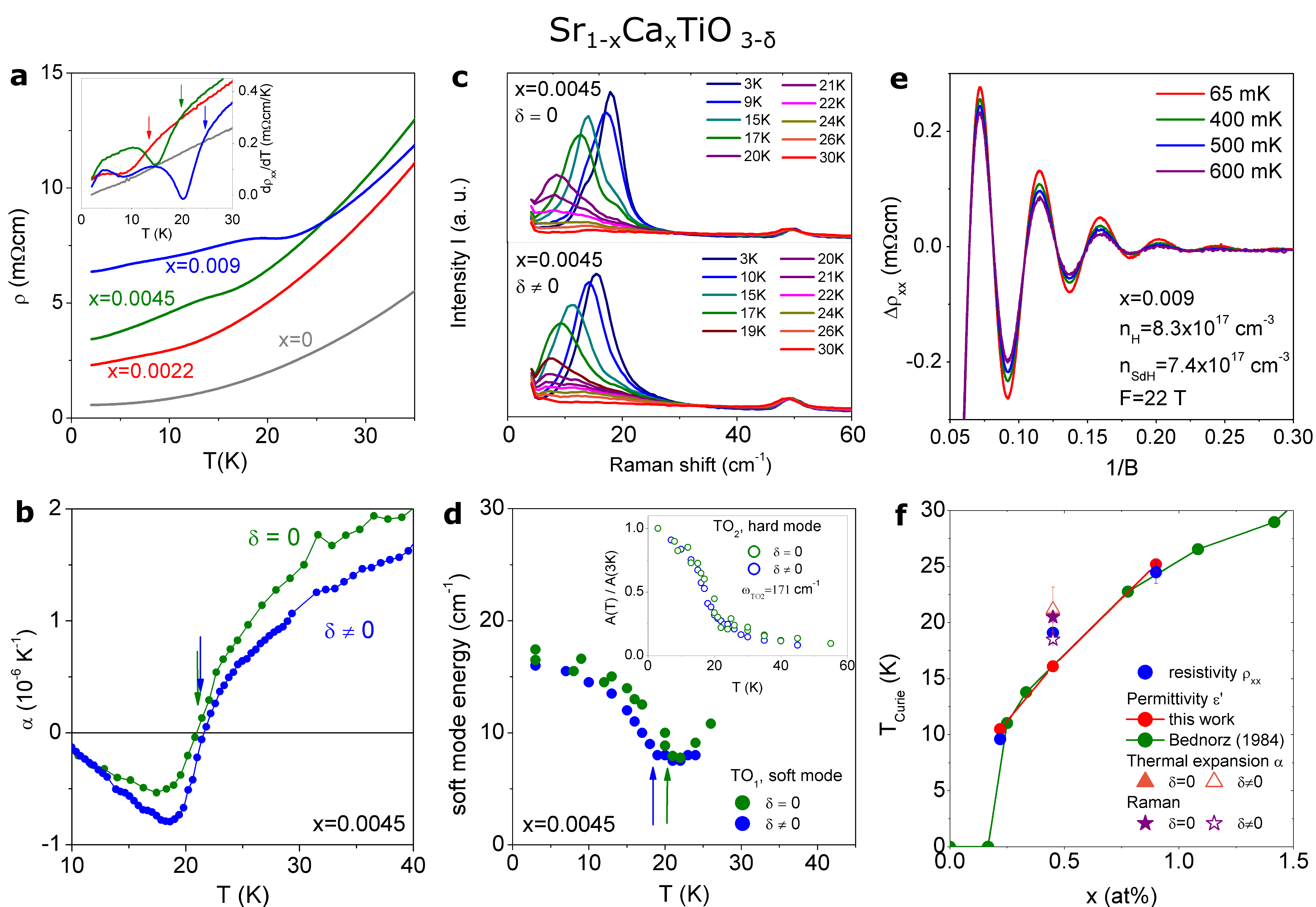}
\caption{\textbf{Coexistence of ferroelectricity with dilute metallicity} \textbf{a)}  Temperature dependence of resistivity in  metallic Sr$_{1-x}$Ca$_{x}$TiO$_{3-\delta}$ ($n\sim 7 \times 10^{17}$ cm$^{-3}$).  Close to the Curie temperature, there is an anomaly, marked by the arrows in the inset showing the temperature dependence of the derivative.   \textbf{b)} Thermal expansion in an insulating and a metallic ($n= 6.6 \times 10^{17}$ cm$^{-3}$) sample. Both show identical anomalies at the same temperature. \textbf{c)} Raman spectrum in (insulating) Sr$_{1-x}$Ca$_{x}$TiO$_{3}$  and (metallic) Sr$_{1-x}$Ca$_{x}$TiO$_{3-\delta}$(n=$9.2\times10^{17}$ cm$^{-3}$). In both cases, the TO$_{1}$ soft mode hardens and grows in intensity upon the entry of the system in the ferroelectric state. \textbf{d}) The hardening of TO$_{1}$ and the change in the integrated intensity of TO$_{2}$ (inset) triggered by the ferroelectric ordering in the two samples. \textbf{e}) Quantum oscillations in  metallic Sr$_{1-x}$Ca$_{x}$TiO$_{3-\delta}$. \textbf{f}) Variation of Curie temperature with Ca content according to different experimental probes. }
\label{Fig2}
\end{figure*}

The first new result of this study is presented in Figure \ref{Fig2}, which shows the persistence of a phase transition structurally identical to the ferroelectric phase transition in dilute metallic Sr$_{1-x}$Ca$_x$TiO$_{3-\delta}$. As seen in Fig. \ref{Fig2}a,  the electric resistivity of these samples shows an anomaly at the Curie temperature of the parent insulating sample (see Fig.\ref{Fig1}b). This observation implies that even in  presence of mobile electrons, the system goes through a ferroelectric-like phase transition at  the same temperature. We checked the presence of this phase transition by two thermodynamic probes. Thermal expansion data is presented in Fig.\ref{Fig2}b. There is a clear anomaly at the Curie temperature. Its magnitude and the temperature at which it occurs are identical in an insulating and a dilute metallic sample with the same calcium content. Our data on sound velocity\cite{Supplement} confirms this. We can therefore safely conclude that the phase transition giving rise to an anomaly in the resistivity of metallic samples (when the carrier concentration is in the range of 10$^{17}$cm$^{-3}$) is structurally identical to the one causing the peak in permittivity in the insulating sample. Further evidence is provided by Raman spectroscopy (Figs. \ref{Fig2}c and d). The entrance to the ferroelectric state is concomitant with the activation of two transverse optical (TO) phonon modes in the Raman spectrum, because of the loss of inversion symmetry. In addition, while the TO$_2$ mode stays hard at 171 cm$^{-1}$, the low energy soft TO$_1$ mode displays a distinctive hardening in the ferroelectric state. As seen in the figure, all these features are present, not only in the insulating sample as reported previously\cite{Kleemann:1997}, but also in a dilute metallic sample. This implies that the low-temperature optical phonon spectrum of the metallic samples does not differ from their insulating ferroelectric counterparts. Note that the anomalies caused by the ferroelectric transition are identical in the metallic and the insulating samples. This means that the presence of mobile electrons in the solid has no incidence on the way the free energy is affected by percolation of electric dipoles. Fig.\ref{Fig2}e shows quantum oscillations of resistivity in dilute Sr$_{1-x}$Ca$_{x}$TiO$_{3-\delta}$. The frequency of the oscillations does not differ from that measured on Ca-free samples at the same carrier concentration \cite{Supplement}, implying that the presence of the ferroelectric-like order neither hinders the connectivity of the Fermi sea, nor modifies its depth.

Our result puts calcium-substituted-oxygen-reduced strontium titanate in the company of a handful of solids known as 'ferroelectric metals'\cite{Shi:2013,Benedek:2016,Kolodiazhnyi:2010}. The expression is used to designate a solid in which mobile electrons are present when a phase transition structurally indistinguishable from a ferroelectric transition occurs. These systems  do not show bulk reversible polarization, the most strict requirement for ferroelectricity. The details of the coexistence between an interconnected Fermi sea and a ferroelectric-like transition (which would have produced a macroscopic polarisation in absence of mobile electrons) remain an open question. However, even at this stage, it is clear that metallicity and ferrolectricity interact with each other in a number of significant ways, as documented below.
\begin{figure*}
\includegraphics[width=0.8\textwidth]{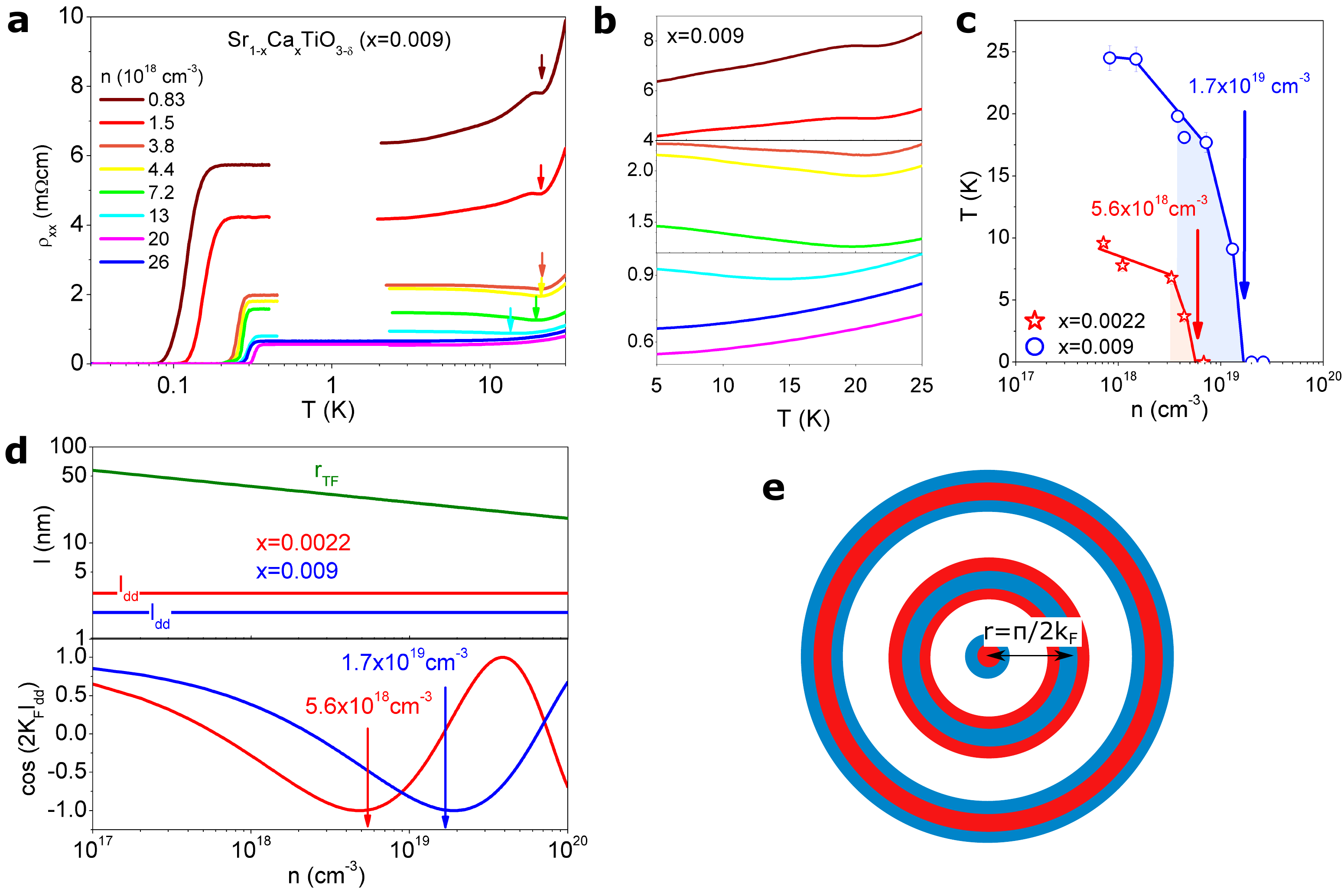}
\caption{\textbf{Evolution of resistivity in Sr$_{1-x}$Ca$_x$TiO$_{3-\delta}$ with increasing carrier concentration} \textbf{a)} Temperature dependence of resistivity in Sr$_{1-x}$Ca$_{x}$TiO$_{3-\delta}$ ($x=0.009$) with different carrier concentrations in a semi-logarithmic plot. Note the presence of two phase transitions. The anomaly in the resistivity marked by an arrow  is induced by the ferroelectric transition. \textbf{b)} Zoom on the same data, showing the emergence of an upturn in resistivity in the intermediate concentration range and the absence of the ferroelectric anomaly (and the recovery of $T$-square resistivity) at higher concentrations. \textbf{c)} The Curie temperature seen by resistivity as a function of carrier concentration. The anomaly disappears above a critical electron concentration, which depends on the calcium content. The shaded area highlights the concentration window at which the system shows a Kondo-like upturn. \textbf{d)} Estimated Thomas-Fermi screening length $r_{TF}$ as a function of carrier density remains much longer than the distance between dipoles, $\ell_{dd}$. On the other hand,  the amplitude of  Friedel oscillation is close  to -1, when the ferroelectric order is destroyed for both Ca concentrations. \textbf{e)} Sketch of Friedel oscillations around a dipole in a Fermi sea. The central circle represents a rattling dipole with red and blue referring to opposite polarities. The dipole will generate concentric circles of oscillations minima/maxima, successively in phase and out-of-phase with the dipole. At a distance r corresponding to $2k_{F}r=\pi$, the oscillations generated by one dipole interferes destructively with a neighboring dipole.}
\label{Fig3}
\end{figure*}

The second  result of this study, a quantum phase transition destroying the ferroelectric-like transition is illustrated in Fig.\ref{Fig3}. Panel a shows the temperature dependence of resistivity in Sr$_{0.991}$Ca$_{0.009}$TiO$_{3-\delta}$. Resistivity shows both a hump at the ferroelectric transition and a drop at the superconducting transition. The samples become more conducting with increasing $\delta$. Zooming on the data in Fig.\ref{Fig3} b, one can see three distinct regimes. At low doping, the percolative ferroelectric phase transition introduces a small but clear anomaly in resistivity. Upon further doping, the resistivity displays an upturn (Fig.\ref{Fig3} b). Finally, above a critical doping, no anomaly in resistivity is detectable and resistivity recovers a purely quadratic temperature dependence as seen in n-doped SrTiO$_{3}$\cite{vandermarel:2011,Lin:2015}. Fig. \ref{Fig3} c shows the temperature of the resistive anomaly as a function of electron concentration for two different Ca contents. The ferroelectric instability is destroyed at $5.8\times10^{18}$  and $1.7\times10^{19}$ cm$^{-3}$  for $x=0.0022$ and $x=0.009$, respectively.

The destruction of the ferroelectric-like order by doping may be ascribed to the screening of direct electric interaction between dipoles by mobile electrons.  This is what has been suggested\cite{Wang:2012} in the case of BaTiO$_{3}$\cite{Kolodiazhnyi:2010}, another 'ferroelectric metal'. In the latter case, the  para- to ferroelectric (cubic-to-tetragonal) phase transition  persists in metallic BaTiO$_{3-\delta}$  up to a critical electron concentration of $1.9 \times 10^{21}$ cm$^{-3}$ \cite{Kolodiazhnyi:2010}. At this carrier density, the Thomas-Fermi screening length, $r_{TF}=\sqrt{\pi a_{B}^{*}/4k_{F}}$, becomes comparable to the lattice constant (and the inter-dipole distance, $\ell_{dd}$) in BaTiO$_{3-\delta}$ (Table 1). As seen in Fig.\ref{Fig3}d, this is not the case of Sr$_{1-x}$Ca$_x$TiO$_{3-\delta}$. For both Ca contents, the destruction of ferroelectricity happens when $r_{TF}$ is more than one order of magnitude longer than $\ell_{dd}$. This motivates us to seek another scenario to explain how metallicity destroys ferroelectricity.

We note that, despite differing in their Curie temperature and dipole concentration by orders of magnitude, ferroelectricity in both Sr$_{1-x}$Ca$_x$TiO$_{3-\delta}$ is destroyed at a threshold of $\sim0.1 e^{-}$/dipole. This leads to an intriguing observation. When the destruction of ferroelectrictity occurs, the interdipole distance, $\ell_{dd}$ and the Fermi wave-vector, k$_{F}$ yield $\cos (2k_{F} l_{dd}) \simeq -1$ (Fig.\ref{Fig3}d). An electric dipole in a Fermi sea would generate Friedel oscillations Fig.\ref{Fig3}e). The interference between oscillations of neighboring dipoles would be destructive in such a condition. We note that, even in the transverse Ising model of impurity-induced ferroelectricity\cite{Wang:1998}, the polarization is not homogeneous and peaks at the Ca-sites,  providing a basis for the dipole-based picture. Further theoretical studies are needed to examine this speculative scenario.

\begin{table*}[htp]
\caption{A comparison of three solids in which the ferroelectric order is destroyed above a critical density of mobile electrons, $n_{c}$. In all three cases, the dipole alignment is destroyed when the product of the interdipole distance, $\ell_{dd}$, and the Fermi wave-vector, $k_{F}$, becomes close to $\pi/2$. This corresponds to a destructive interference between Friedel oscillations of neighboring dipoles.}
\centering 
\begin{tabular}{|c| c| c |c| c |c |c|} 
\hline\hline 
Compound &$ n_c [cm^{-3}]$ &$ \ell_{dd} [nm]$ & $r_{TF}(n=n_c)[nm]$ & k$_{F} (n=n_c)[nm^{-1}]$ & $r_{TF}/\ell_{dd}$ & k$_{F}\ell_{dd}$ \\ [0.5ex] 
\hline 

BaTiO$_{3-\delta}$ \cite{Kolodiazhnyi:2010,Wang:2012}& 1.9 10$^{21}$ &0.4 & 0.4 & 3.82 & 1& 1.53\\ \hline
Sr$_{0.991}$Ca$_{0.009}$TiO$_{3-\delta}$& 1.7 10$^{19}$ & 1.9 & 24 & 0.79 & 12.6& 1.50 \\\hline
Sr$_{0.998}$Ca$_{0.002}$TiO$_{3-\delta}$& 5.6 10$^{18}$ & 3.0 & 29 & 0.55 & 9.7& 1.65 \\ [1ex] 
\hline 
\hline 
\end{tabular}
\label{table:1} 
\end{table*}

Further signature of coupling between dipoles and mobile electrons is the upturn in resistivity at intermediate carrier concentrations (Fig.\ref{Fig3} b). The Kondo effect, a many-body resonant scattering due to coupling between  the Fermi sea and an alien spin\cite{Hewson:1993} can give rise to such an upturn. Any quantum degeneracy of a localized state (and not only spin) may produce an Abrikosov-Suhl resonance and a variety of non-magnetic counterparts of the Kondo effect have been experimentally observed. In  PbTe, another dilute metal close to a ferroelectric instability, the introduction of Tl dopants leads to (dilute superconductivity and) an upturn in resistivity, attributed to the charge Kondo effect \cite{Matsushita:2005}. A dipolar Kondo effect is a reasonable candidate for explaining the upturn in resistivity seen here.  As seen, in Fig.\ref{Fig3} c, it only occurs when percolation of the ferroelectric droplets is rapidly degrading with increasing carrier density, providing constraints for any scenario based on Kondo resonance occurring near the quantum phase transition.

\begin{figure*}
\includegraphics[width=0.8\textwidth]{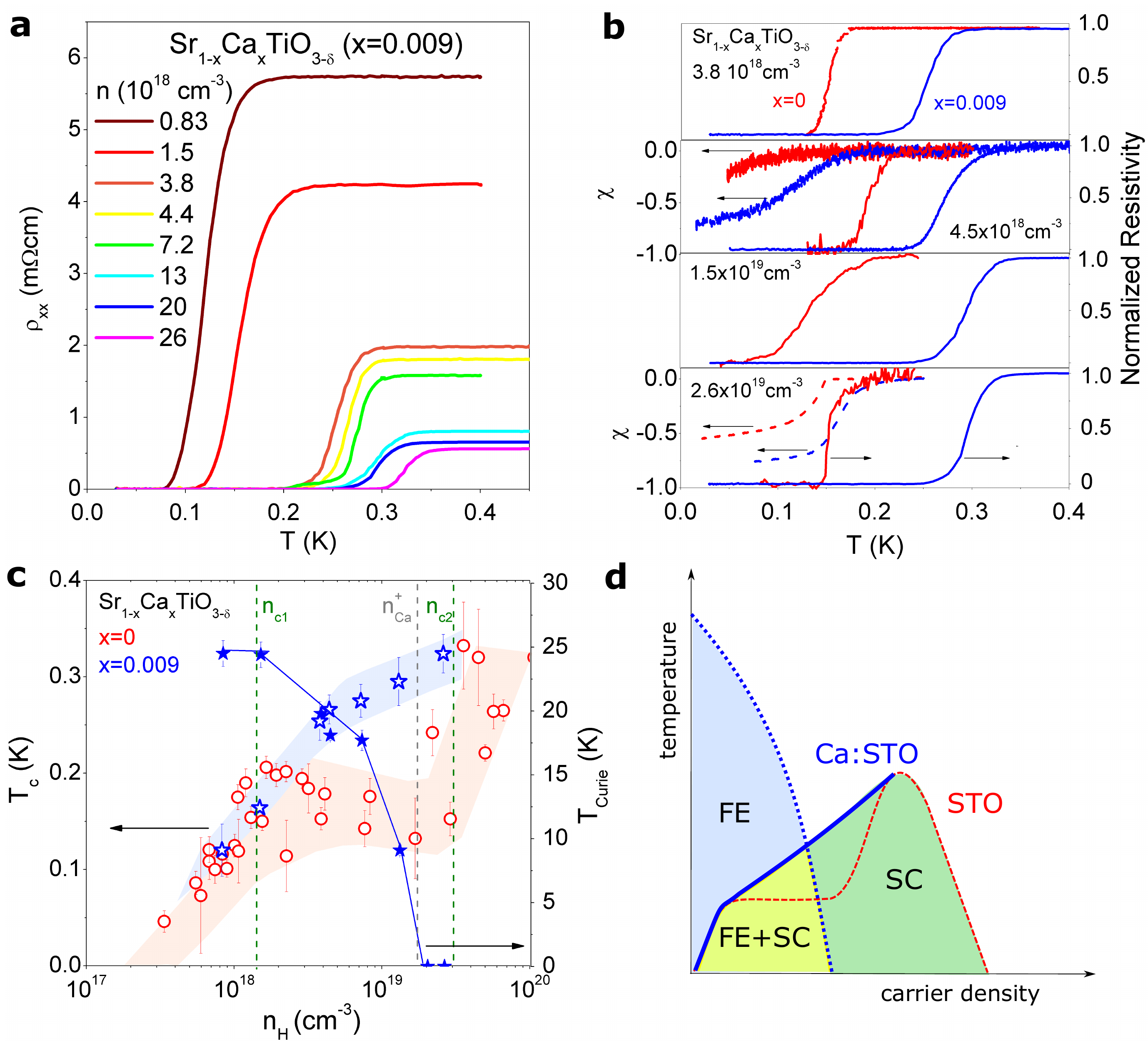}
\caption{\textbf{Evolution of the superconducting transition temperature} \textbf{a)} With increasing doping, the magnitude of normal-state resistivity decreases and the critical temperature rises. \textbf{b)} Superconducting transition seen by resistivity and ac susceptibility in Ca-substituted and Ca-free samples with comparable carrier concentration in the range where Ca substitution enhances T$_{c}$. \textbf{c)} Resistive critical temperature, $T_c$, as a function of carrier concentration in SrTiO$_{3-\delta}$ and  Sr$_{0.991}$Ca$_{0.009}$TiO$_{3-\delta}$. Additional Fermi surface pockets emerge at $n_{c1}$ and $n_{c2}$\cite{Lin:2014}. The variation of Curie temperature with carrier concentration is also shown. Note the increase in the superconducting Tc is enhanced in the vicinity of n$^{*}_{Ca}$, the concentration at which the ferroelectric-like order is destroyed. \textbf{d)} Sketch of the phase diagram showing the region where superconducting (SC) and ferroelectric (FE) order coexist and the increase in T$_{c}$ induced by Ca substitution.}
\label{Fig4}
\end{figure*}

Fig.\ref{Fig4} shows how the superconducting phase diagram is affected by calcium doping in Sr$_{0.991}$Ca$_{0.009}$TiO$_{3-\delta}$.  As seen in panel a, with increasing carrier concentration the superconducting transition  steadily shifts to higher temperatures. At a given concentration, it occurs at a higher temperature in Ca-substituted samples compared to Ca-free ones (panel b).  Note that the critical temperature according to bulk probes (specific heat, thermal conductivity and magnetic susceptibility) is significantly lower than the onset of the resistive transition\cite{Lin:2014b}. As seen in Fig.\ref{Fig4}c, the onset of superconductivity seen by AC-susceptibility, which sets in below the resistive T$_{c}$, shifts to a higher temperature after Ca substitution. According to our data, at least in a finite window between the two local maxima close to two critical doping levels, $n_{c1}$ and $n_{c2}$, of Ca-free SrTiO$_{3-\delta}$\cite{Lin:2014}, superconductivity is strengthened by Ca substitution. As sketched in panel d, there is a region in the phase diagram in which superconductivity and ferroelectricity coexist. Since one cannot separate the orbital and spin components of the order parameter in a non-centrosymmetric superconductor\cite{Bauer:2012}, this region would be a very appealing playground for future research.

The enhancement of critical temperature by Ca substitution is observed in the vicinity of the quantum phase transition leading to the destruction of ferroelectricity. This provides a new input for the ongoing debate on the microscopic origin of superconductivity\cite{Rowley:2014,Edge:2015, Gorkov2015,Ruhman2016}. Calcium substitution leads to a higher dielectric constant at low temperature (Fig.\ref{Fig1} b), which would screen Coulomb repulsion between electrons. It hardens the soft ferroelectic mode(Fig.\ref{Fig2} b). Both these features may be relevant to the observed enhancement in superconducting T$_{c}$. Edge et \emph{al.}\cite{Edge:2015} have recently proposed that substituting O$^{16}$ with O$^{18}$ should drastically change the superconducting dome of n-doped SrTiO$_{3}$. They predicted that such a substitution would enhance the highest critical temperature and shift it to lower doping. This is a quantum critical ferroelectric\cite{Rowley:2014} scenario, in which the maximum T$_{c}$ is pinned to the destruction of the ferroelectric order. Our observation is in qualitative agreement with this scenario.  We find that T$_{c}$ is enhanced in the vicinity of the critical doping at which the ferroelectric-like order is destroyed (n$^{*}_{Ca} =1.7\times 10^{19}$cm$^{-3}$ in Sr$_{0.991}$Ca$_{0.009}$TiO$_{3-\delta}$). For this level of Ca content, n$^{*}_{Ca}$ lies between the two critical doping levels, $n_{c1}$ and $n_{c2}$, of  Ca-free SrTiO$_{3-\delta}$ where the T$_{c}$ is almost flat\cite{Lin:2014}.  A definite confirmation of the quantum critical scenario requires an extensive study employing  bulk probes\cite{Lin:2014b} and confirming that the enhancement in critical temperature occurs always near n$^{*}_{Ca}$, which shifts with Ca content, and is not pinned to $n_{c1}$ and $n_{c2}$.

This work has been supported  by ANR (through the SUPERFIELD and QUANTUM LIMIT projects), by an \emph{Ile de France} regional grant, by Fonds ESPCI-Paris, by DFG research grant HE-3219/2-1 and by the Institutional Strategy of the University of Cologne within the German Excellence Initiative. XL is supported by the Alexander von Humboldt Foundation.

\clearpage

\begin{widetext}

\section*{Supplementary Information}

\section{\label{SM}Samples and experimental techniques}
For this work we used commercially obtained SrTiO$_{3}$ and Sr$_{1-x}$Ca$_{x}$TiO$_{3}$ ($x=0.0022$, 0.0045 and 0.009) single crystals. The nominal calcium concentration of two samples was checked using the Secondary Ion Mass Spectrometry (SIMS) analysis technique as detailed previously\cite{DeLima:2015}. The oxygen content has been changed by heating the samples in vacuum (pressure $10^{-6}-10^{-7}$ mbar) to temperatures of $775-1100$ $^\circ$C. In order to attain carrier densities above $\sim 4\times 10^{18}$ cm$^{-3}$, a piece of titanium has been placed next to the sample during heating. Ohmic contacts have been realized prior to oxygen removal by evaporation of gold contact pads.\\
The electrical measurements have been performed in a Quantum Design Physical Property Measurement System (PPMS) between 1.8 and 300 K as well as in a 17 T dilution refrigerator with a base temperature of 26 mK. Detailed electrical transport information on all samples presented in the main text is listed in Tab. S\ref{tab:Samples} of this supplement.\\
The ac magnetic susceptibility was measured in a homemade set-up, comprising a primary coil and a compensating pick-up coil with two sub-coils with their turns in opposite direction. A Lock-in amplifier was utilized to supply the exciting ac current and pick up the induced voltage signal. The applied ac field was as low as 10 mG with a frequency of 16 kHz.\\
The dielectric permittivity measurements were performed employing a frequency-response analyzer ({\sc Novocontrol} Alpha-Analyzer). Using silver paint, the plate-like samples were prepared as capacitors with typical electrode dimensions of $3 \times 3$~mm$^2$ and a typical thickness of 0.5-0.85~mm. For the evaluation of the as-measured data $C_{meas}$, passivated surface layers were assumed, as described in \cite{Aso:1976}. Such layers can be considered as additional capacitors $C_{surf}$ in series to the remaining bulk specimen $C_{bulk}$, which limits the total capacitance data. Therefore the data were corrected assuming a temperature independent surface contribution $C_{bulk}{-1}=C_{meas}{-1}-C_{surf}{-1}$, which results in $\varepsilon(T)$ curves comparable to literature data on surface-etched samples \cite{Aso:1976}. Measurements of $P(E)$-hysteresis loops were performed using the same setup with an additional high-voltage module ({\sc Novocontrol} HVB1000). The actual field dependent polarization was calculated from the non-linear dielectric permittivities up to the tenth order as described in \cite{Niermann:2014}. The thermo-remanent polarization data was gained from the integrated pyro-current as collected with an electrometer (Keithley 6517) after cooling in a poling field of approximately 120~V/mm.\\
A home-built capacitance dilatometer has been used to detect the uniaxial length changes $\Delta L(T)$ while continuously heating the crystal from about 5 to 150~K with a rate of about $0.1$~K/min. Here, $\Delta L(T)$ was measured along the [100] directions of Sr$_{1-x}$Ca$_{x}$TiO$_{3-\delta}$ single crystals with total lengths $L_0\simeq 2$~mm and the uniaxial thermal expansion coefficient $\alpha=1/L_0 \partial \Delta L/\partial T$ has been derived numerically.\\
The Raman measurements were performed using the 532 nm line of Diode Pumped Solid State (DPSS) laser. An incident power of 5mW was focused on a spot of dimension 50 $\times$ 80 $\mu$m approximately. Power dependance measurements at low temperature indicated negligible laser heating for this incident power. The inelastically scattered photons were analyzed using a triple grating spectrometer working in subtractive configuration and equipped with a nitrogen cooled Charge Coupled Device (CCD) camera. The spectral resolution was about 1.5 cm$^{-1}$. All spectra were recorded with linearly polarized and parallel incoming and outgoing photons. The crystals were cooled using a close-cycle optical cryostat with a base temperature of 3 K.\\

\begin{table}
\centering
\caption{\label{tab:Samples} Details of the Sr$_{1-x}$Ca$_{x}$TiO$_{3-\delta}$ samples. Hall carrier density ($n_H$), room temperature ($\rho_{\textnormal{300K}}$) and 2 K ($\rho_{\textnormal{2K}}$) resistivity, the ratio RRR$=\rho_{\textnormal{300K}}/\rho_{\textnormal{2K}}$, the Hall mobility at 2 K ($\mu_{H-\textnormal{2K}}$) and the Curie temperature seen by resistivity are specified.}
\footnotesize
\begin{tabular}{ccccccc}
\hline
$x$ 		& $n_H$  			& $\rho_{\textnormal{300K}}$ & $\rho_{\textnormal{2K}}$ 	& RRR & $\mu_{H-\textnormal{2K}}$ & $T_{Curie,{\rho}_{xx}}$ \\
	& $10^{18}$cm$^{-3}$  	& m$\Omega$cm 				& m$\Omega$cm  							& 		& cm$^2$/Vs & K\\
\hline
\hline
	0.0022 & 0.72 & 1840 & 2.30 & 800 & 3770 & $9.6 \pm 0.4$ \\
	0.0022 & 1.1 & 1380 & 2.31 & 600 & 2456 & $ 7.8\pm 0.5$ \\
	0.0022 & 3.3 & 553 & 1.08 & 512 & 1751  & $ 6.8\pm 0.5$ \\
	0.0022 & 4.4 & 344 & 0.847 & 406 & 1675  & $ 3.7\pm 0.4 $\\
	0.0022 & 6.8 & 87 & 0.625 & 139 & 1469 &  0\\
	\hline
	0.0045 & 0.66 & 1830 & 3.43 & 535 & 2760 & $ 19.1\pm 1 $ \\
	\hline
	0.009 & 0.83 & 1470 & 6.37 & 231 & 1183 & $24.5\pm 1$\\
	0.009 & 1.5 & 857 &	4.17 & 206 & 1005 & $24.4 \pm 1$\\
	0.009 & 3.8 & 394 & 2.27 & 174 & 724 & $19.8 \pm 0.4$\\
	0.009 & 4.4 & 346 & 2.17 & 159 & 654 & $ 18.1 \pm 0.4$\\
	0.009 & 7.2 & 217 & 1.48 & 147 & 586 & $17.7 \pm 0.8$\\
	0.009 & 13 & 132 &	0.941 & 140 & 510 & $9.1 \pm 0.4$\\
	0.009 & 20 & 91.3 & 0.656 & 139 & 476 & 0\\
	0.009 & 26 & 70.4 & 0.542 & 130 & 443 & 0\\
	\hline
\end{tabular}
\normalsize
\end{table}

\section{\label{Tdep}Temperature dependence of resistivity}
Figures S\ref{FigSI1} and S\ref{FigSI2} plot the resistivity $\rho_{xx}$ as well as its derivative $d\rho_{xx}/dT$ measured on Sr$_{1-x}$Ca$_{x}$TiO$_{3-\delta}$ samples with Ca contents of $x=0.0022$ and 0.009, respectively, as a function of temperature $T$. The arrows mark the temperatures associated with the resistivity anomaly and the Curie temperature $T_{Curie,\rho_{xx}}$ seen in resistivity plotted in Fig. 2c of the main text (see also Tab. S\ref{tab:Samples}). The temperatures $T_{Curie,\rho_{xx}}$ have been taken as the temperature at which $d\rho_{xx}/dT$ shows a kink.\\
\begin{figure}
\centering
\includegraphics[width=0.6\textwidth]{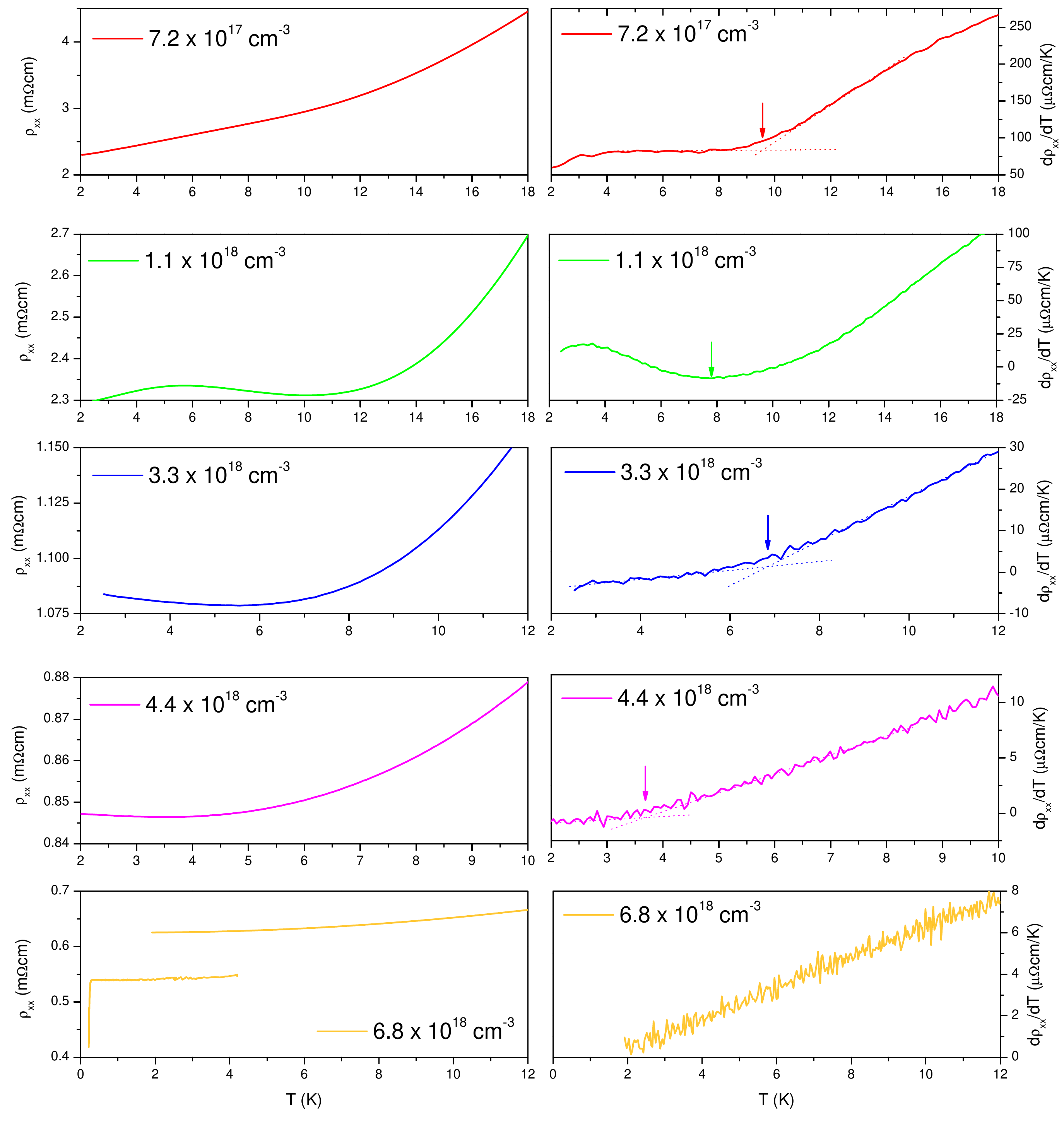}
\caption{Temperature dependence of resistivity $\rho_{xx}$ and its first derivative $d\rho_{xx}/dT$ in metallic Sr$_{0.9978}$Ca$_{0.0022}$TiO$_{3-\delta}$. The arrows mark the temperatures associated with the resistivity anomaly and the Curie temperature $T_{Curie,\rho_{xx}}$ seen in resistivity.}
\label{FigSI1}
\end{figure}

\begin{figure}
\centering
\includegraphics[width=0.6\textwidth]{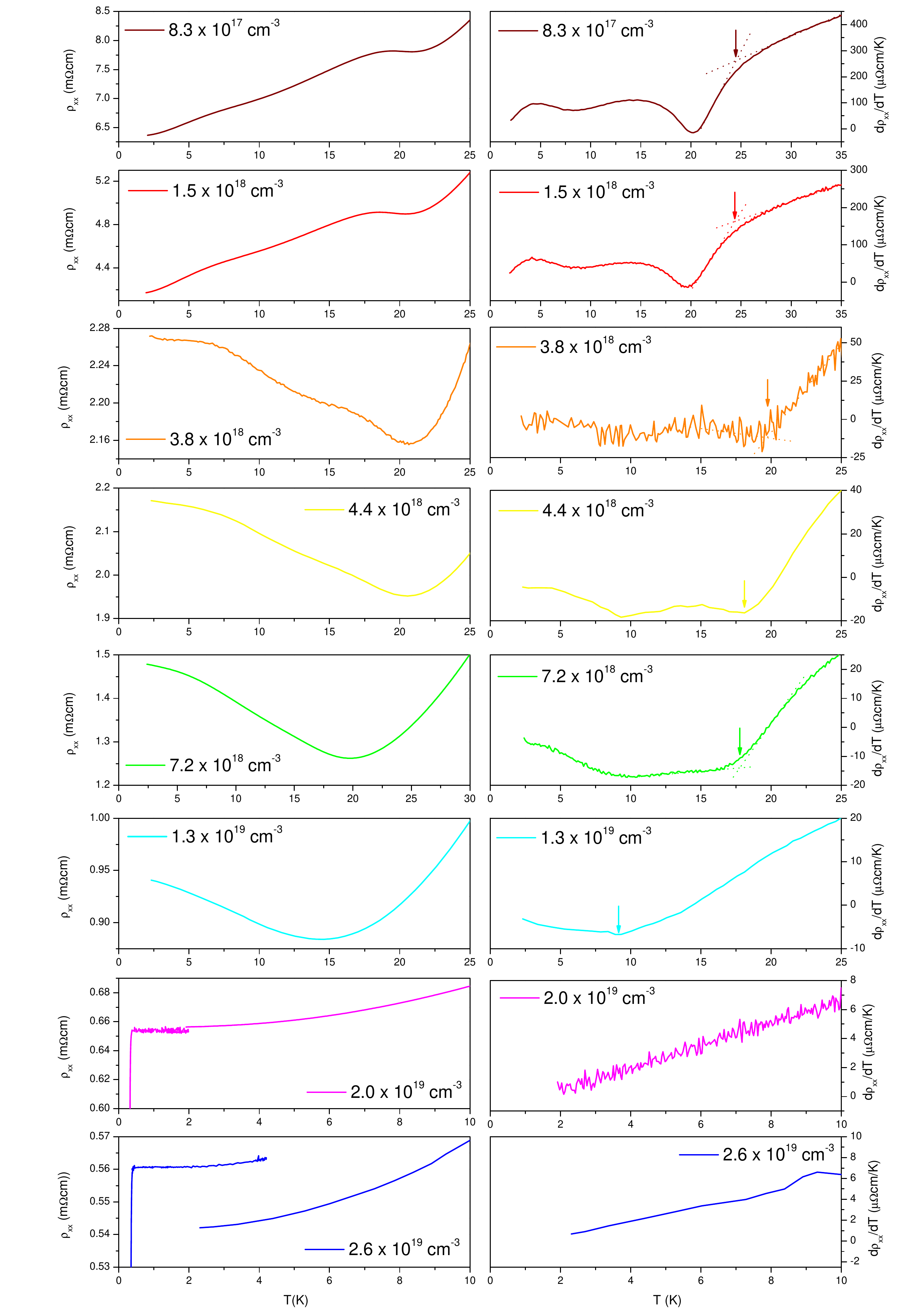}
\caption{Temperature dependence of resistivity $\rho_{xx}$ and its first derivative $d\rho_{xx}/dT$ in metallic Sr$_{0.991}$Ca$_{0.009}$TiO$_{3-\delta}$.}
\label{FigSI2}
\end{figure}

\clearpage

\section{\label{US}Temperature dependence of thermo-remanent polarization}
Fig. S\ref{FigSI5tp} displays the thermo-remanent polarization in the system Sr$_{0.991}$Ca$_{0.009}$TiO$_{3}$. A  theoretical description of the quantum ferroelectric regime according to the transverse Ising model yields values for the saturation polarization reaching up to 20\,mC/m$^2$ for a comparable Ca concentration \cite{Guo:2012}. However, as we are dealing only with the remanant polarization, the maximum value at low temperatures lies around 20\,mC/m$^2$, which is smaller due to domain formation but is still of the same order of magnitude. Upon heating above the ferroelectric ordering temperature $T_{Curie}$ near 25\,K $P(T)$ shows a steep fall, but does not vanish completely. Above $T_{Curie}$, there is a finite frozen polarization, as seen by the opening of the $P(E)$ hysteresis loops  shown in the inset of Fig. S\ref{FigSI5tp}. Obviously, polar entities like ferroelectric clusters or polar structural domain walls persist to temperatures well above the ferroelectric transition. The latter is in accord to results gained from ultrasound experiments even in pure STO \cite{Salje:2013}. A corresponding characterization of the oxygen-reduced samples obviously cannot be made as the macroscopic polarization features are shielded by the metallic background.
\begin{figure}[htbp]
\centering
\includegraphics[width=0.4\textwidth]{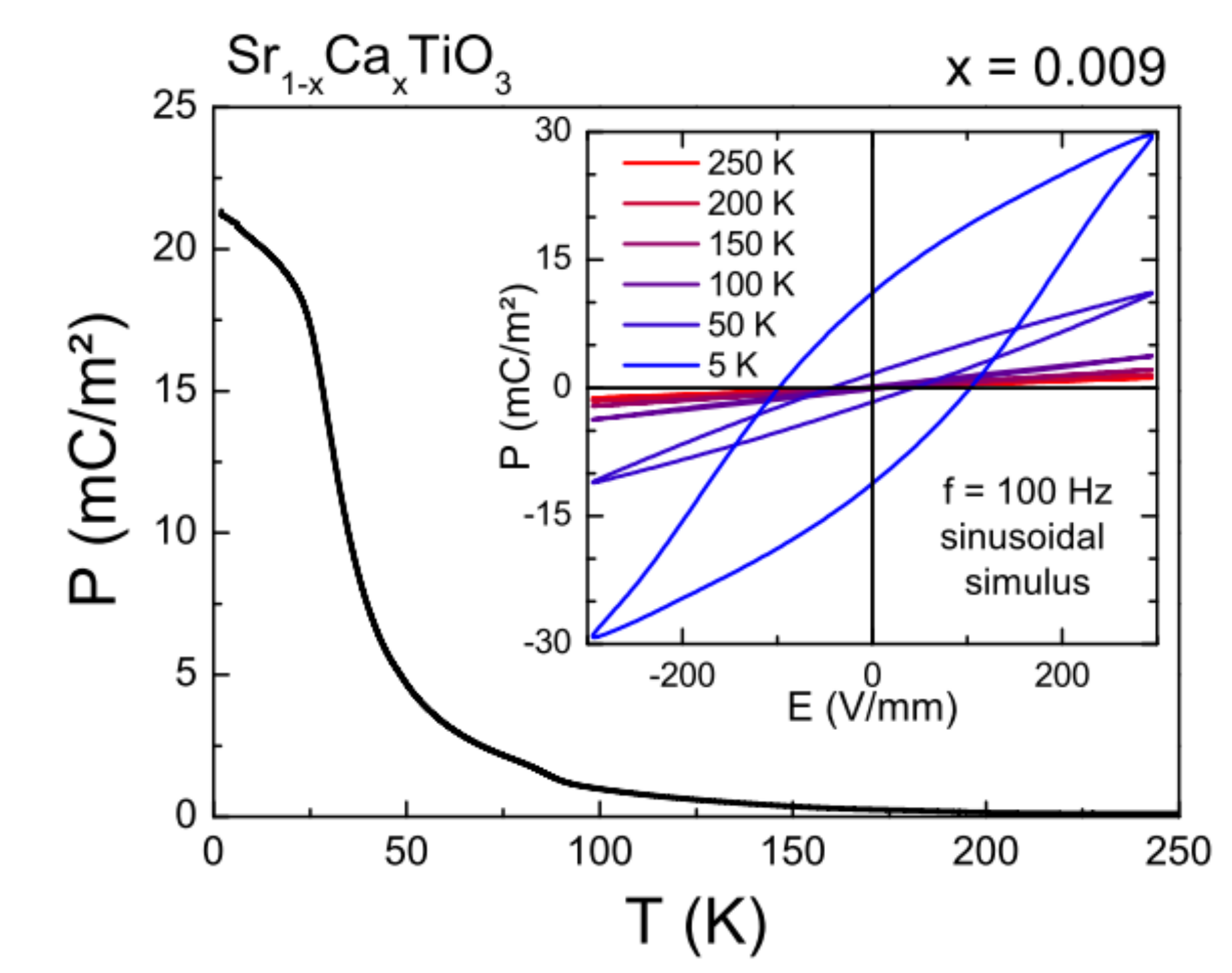}
\caption{Thermo-remanent polarization measured in an insulating sample of Sr$_{0.991}$Ca$_{0.009}$TiO$_{3}$ after cooling in a poling field of 120\,V/mm. The inset shows $P(E)$ loops measured at various temperatures between 5 and 250\,K.}
\label{FigSI5tp}
\end{figure}

\section{\label{US}Detection of the phase transition with sound-velocity measurements}
The sound velocity was measured in transmission geometry using longitudinal 10~MHz PZT-transducers as emitter and detector. A network analyzer (Rohde \& Schwarz ZVB4) with time domain option was used to determine the transit time trough plate-like samples with a length of typically 5 mm\cite{Balashova:1996}. The FFT-representation of the response carries an absolute time resolution of the reciprocal resonance frequency, i.e., approximately 100 nS. However, relative changes of the transit time can be determined with much higher resolution.\\
The sound velocity as a function of temperature in an insulating ($\delta = 0$) and metallic ($\delta \neq 0$) Sr$_{0.991}$Ca$_{0.009}$TiO$_{3-\delta}$ sample is shown in Fig. S\ref{FigSI7}. In both cases the phase transition gives rise to an anomaly near the Curie temperature.
\begin{figure}
\centering
\includegraphics[width=0.4\textwidth]{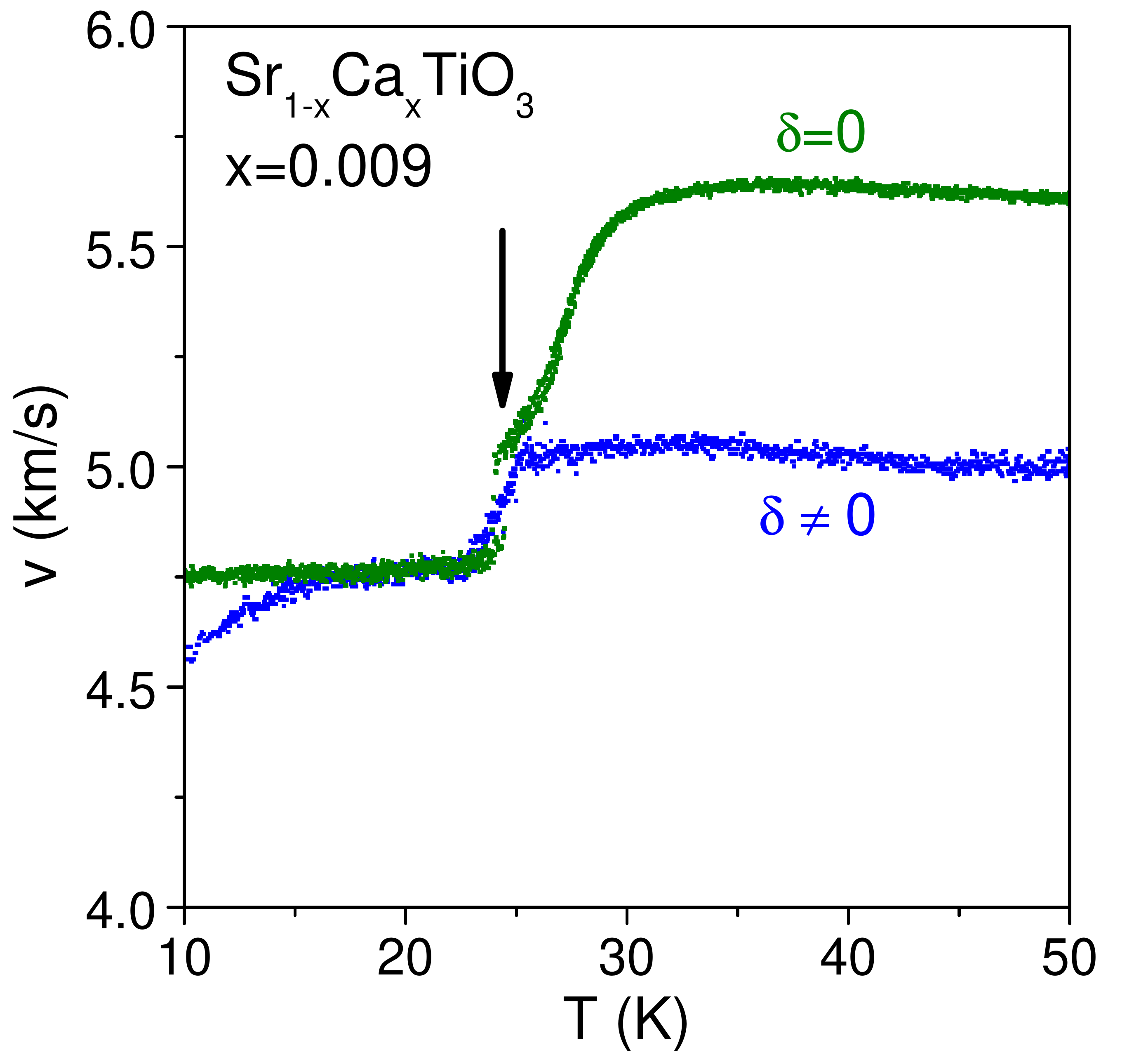}
\caption{Sound velocity as a function of temperature in insulating ($\delta = 0$) and metallic ($\delta \neq 0$, $n=1.4 \times 10^{18}$ cm$^{-3}$) Sr$_{0.991}$Ca$_{0.009}$TiO$_{3-\delta}$ samples. The arrow depicts the Curie temperature.}
\label{FigSI7}
\end{figure}

\section{\label{QO}Quantum oscillations}

\begin{figure}
\centering
\includegraphics[width=0.4\textwidth]{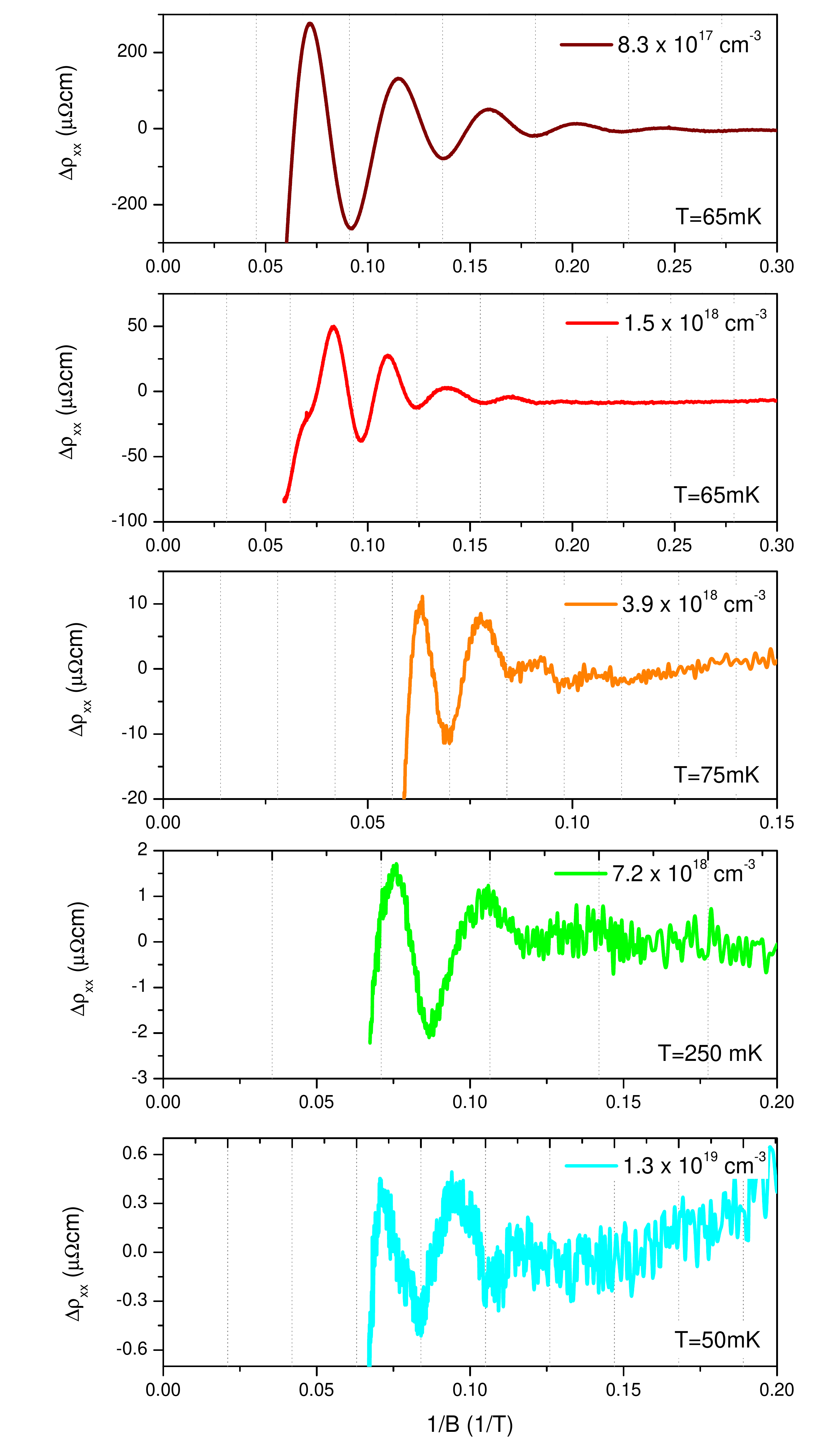}
\caption{Quantum oscillations detected at low temperatures on metallic Sr$_{0.991}$Ca$_{0.009}$TiO$_{3-\delta}$. Each panel plots the oscillating part $\Delta\rho_{xx}$ of the magnetoresistance, obtained after subtracting a smooth background, as a function of inverse magnetic field $1/B$.}
\label{FigSI3}
\end{figure}

\begin{figure}
\centering
\includegraphics[width=0.4\textwidth]{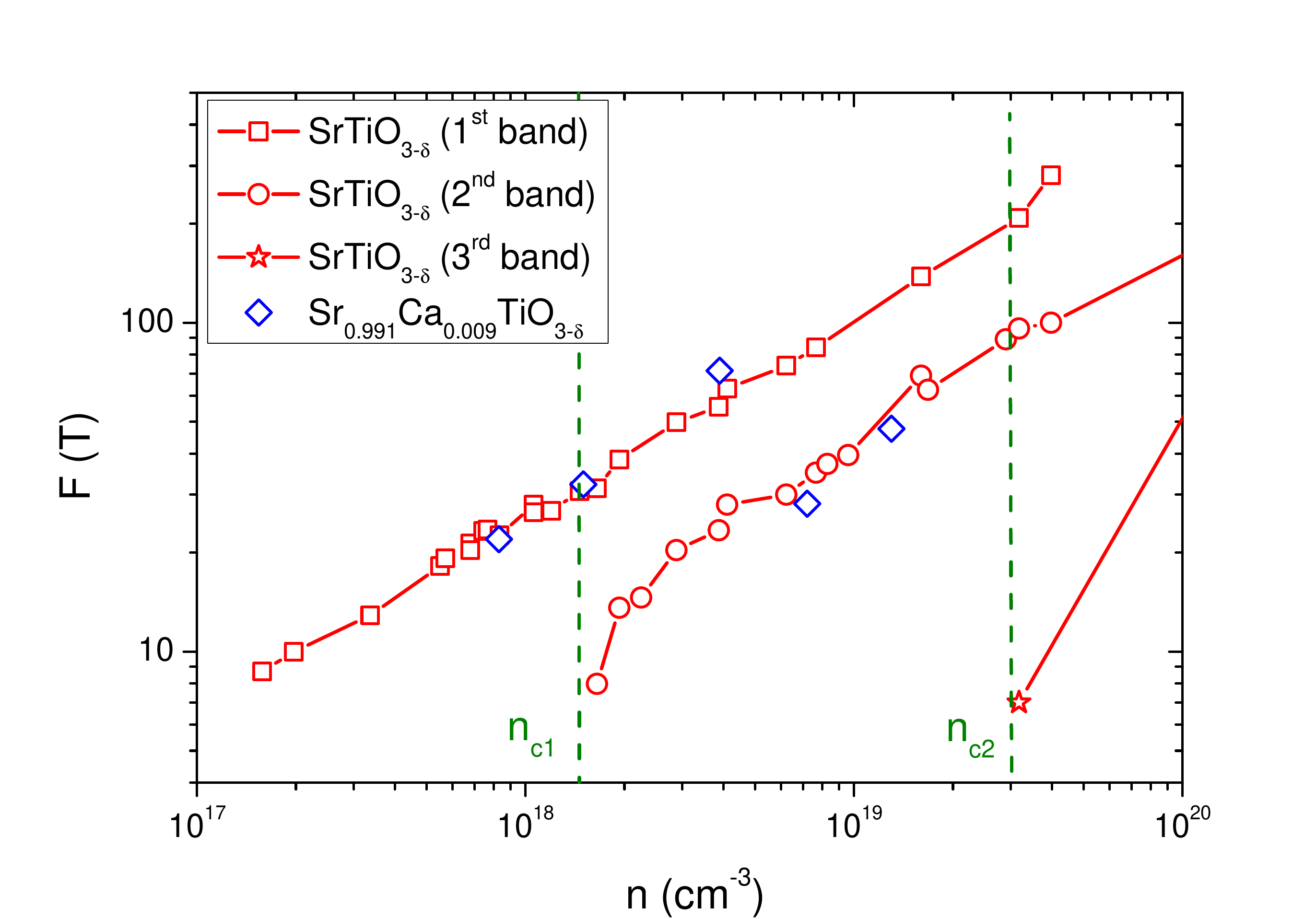}
\caption{Oscillation period F detected on Sr$_{0.991}$Ca$_{0.009}$TiO$_{3-\delta}$ as a function of carrier density together with data obtained on SrTiO$_3$ \cite{Lin:2014}. The doping levels $n_{c1}$ and $n_{c2}$ correspond to the threshold for the filling of a second and third band.}
\label{FigSI4}
\end{figure}
Quantum oscillations in SrTiO$_{3-\delta}$ have been systematically studied for a large range of carrier densities ($10^{17} - 10^{20}$ cm$^{-3}$) using electric and thermoelectric measurements \cite{Lin:2013,Lin:2014}. Fig. S\ref{FigSI3} plots the oscillating part $\Delta\rho_{xx}$ of the magnetoresistance, obtained after subtracting a smooth background, as a function of inverse magnetic field $1/B$ for metallic Sr$_{1-x}$Ca$_{x}$TiO$_{3-\delta}$ ($x=0.0091$) samples. The detected oscillation periods $F$ are plotted as a function of carrier density in Fig. S\ref{FigSI4} and compared with data obtained on SrTiO$_{3-\delta}$ \cite{Lin:2014}. Lin et al. identified two critical doping levels $n_{c1}\approx 1.5\times 10^{18}$ and $n_{c2}\approx 3 \times 10^{19}$ cm$^{-3}$ that correspond to the threshold for the filling of a second and third band and which are associated with the appearance of a second or third oscillation period, respectively. As seen in Fig. S\ref{FigSI4}, the detected periods obtained on Sr$_{1-x}$Ca$_{x}$TiO$_{3-\delta}$ agree well with those obtained on Ca-free STO. However, due to the $3-5$ times lower mobility in the Ca-doped samples compared to pure STO, only the lowest oscillation period could be resolved with certainty for $n>n_{c1}$.\\
Even for low carrier concentrations below $n_{c1}$ the Fermi surface of n-doped SrTiO$_{3-\delta}$ is not a perfect sphere but an ellipsoid squeezed along the c-axis due to the tetragonal distortion of the lattice. As done in \cite{Lin:2014} we estimated the carrier concentration $n_{SdH}$ from the oscillations using the magnitude of the tetragonal distortion reported by Allen et al. \cite{Allen:2013}. The values of $n_{SdH}$ given in the main text were calculated as $n_{SdH}=1.26\times n_{sphere}$ with $n_{sphere}=k_{F}^{3}/3\pi^2$ and $k_{F}=\sqrt{2eF/\hbar}$ with the oscillation period $F$.

\end{widetext}


\begin{thebibliography}{XX}

\bibitem{Muller:1979}
K. A. M\"uller and H. Burkhard. SrTiO$_{3}$: An intrinsic quantum paraelectric below 4 K. Phys. Rev. B \textbf{19}, 3593 (1979).

\bibitem{Schooley:1964}
J. F. Schooley, W. R. Hosler, and M. L. Cohen. Superconductivity in Semiconducting SrTiO$_{3}$.  Phys. Rev. Lett. \textbf{12}, 474 (1964).

\bibitem{Bednorz:1984}
J. G. Bednorz and K. A. M\"uller. Sr$_{1-x}$Ca$_{x}$TiO$_{3}$ : An XY Quantum Ferroelectric with Transition to Randomness. Phys. Rev. Lett. \textbf{52}, 2289 (1984).

\bibitem{Rowley:2014}
S. E. Rowley, L. J. Spalek, R. P. Smith, M. P. M. Dean, M. Itoh, J. F. Scott, G. G. Lonzarich and S. S. Saxena. Ferroelectric quantum criticality.  Nature Phys. \textbf{10}, 367 (2014).

\bibitem{Koonce:1967}
C. S. Koonce, M. L. Cohen, J. F. Schooley, W. R. Hosler and E. R. Pfeiffer.  Superconducting Transition Temperatures of Semiconducting SrTiO$_{3}$. \emph{Phys. Rev.} \textbf{163}, 380 (1967)

\bibitem{Lin:2013}
X. Lin, Z. Zhu, B. Fauqu\'{e}, and K. Behnia. Fermi Surface of the Most Dilute Superconductor. Phys. Rev. X \textbf{3}, 021002 (2013).

\bibitem{Lin:2014}
X. Lin, G. Bridoux, A. Gourgout, G. Seyfarth, S. Kr\"{a}mer, M. Nardone, B. Fauqu\'e and K. Behnia. Critical Doping for the Onset of a Two-Band Superconducting Ground State in  SrTiO$_{3- \delta}$. Phys. Rev. Lett. \textbf{112}, 207002 (2014).

\bibitem{Takada}
Y. Takada. Theory of superconductivity in polar semiconductors and its application to n-type semiconducting SrTiO$_{3}$, J. Phys. Soc. Jpn. \textbf{49} 1267(1980).

\bibitem{Ruhman2016}
J. Ruhman and P. A. Lee. Superconductivity at very low density: the case of strontium titanate. Phys. Rev. B \textbf{94}, 224515 (2016).

\bibitem{Edge:2015}
J. M. Edge, Y. Kedem, U. Aschauer, N. A. Spaldin, and A. V. Balatsky. Quantum Critical Origin of the Superconducting Dome in  SrTiO$_{3}$, Phys. Rev. Lett. \textbf{115}, 247002 (2015).

\bibitem{Gorkov2015}
L. P. Gorkov. Phonon mechanism in the most dilute superconductor: n-type SrTiO$_{3}$. PNAS \textbf{113}, 4646 (2016).

\bibitem{Itoh:1999}
M. Itoh, R. Wang, Y. Inaguma, T. Yamaguchi, Y-J. Shan, and T. Nakamura. Ferroelectricity Induced by Oxygen Isotope Exchange in Strontium Titanate Perovskite. Phys. Rev. Lett. \textbf{82}, 3540 (1999).

\bibitem{Uwe:1976}
H. Uwe and T. Sakudo. Stress-induced ferroelectricity and soft phonon modes in SrTiO$_{3}$. Phys. Rev. B \textbf{13}, 271 (1976).

\bibitem{Kleemann:1988}
W. Kleemann, F. J. Sch\"{a}fer, K. A. M\"{u}ller and J. G. Bednorz. Domain state properties of the random-field xy-model system Sr$_{1-x}$Ca$_x$TiO$_{3}$. Ferroelectrics \textbf{80}, 297 (1988).

\bibitem{Bianchi:1994}
U. Bianchi, W. Kleemann and J. G. Bednorz. Raman scattering of ferroelectric Sr$_{1-x}$Ca$_{x}$TiO$_{3}$, x=0.007. J. Phys.: Condens. Matter \textbf{6}, 1229 (1994).

\bibitem{Kleemann:1997}
W. Kleemann, A. Albertini, M. Kuss and R. Lindner. Optical detection of symmetry breaking on a nanoscale in SrTiO$_{3}$:Ca. Ferroelectrics \textbf{203}, 57 (1997).

\bibitem{Kleemann:2000}
W. Kleemann, J Dec, Y. G. Wang, P. Lehnen and S.A Prosandeev. Phase transitions and relaxor properties of doped quantum paraelectrics. J.  Phys.  Chem.  Sol. \textbf{61},  167 (2000).

\bibitem{Hemberger:1996}
J. Hemberger, M. Nicklas, R. Viana, P. Lunkenheimer, A. Loidl and R. B\"{o}hmer. Quantum paraelectric and induced ferroelectric states in SrTiO$_{3}$. J. Phys.: Condens. Matter \textbf{8}, 4673 (1996).

\bibitem{Carpenter:2006}
M. A. Carpenter, C. J Howard, K. S. Knight and Z. Zhang. Structural relationships and a phase diagram for (Ca,Sr)TiO$_{3}$ perovskites. J. Phys. Condens. Matter \textbf{18}, 10725 (2006).

\bibitem{Ang:2001}
C. Ang, A. S. Bhalla, and L. E. Cross. Dielectric behavior of paraelectric KTaO$_{3}$ , CaTiO$_{3}$, and Ln$_{1/2}$Na$_{1/2}$TiO$_{3}$ under a dc electric field. \emph{Phys. Rev. B} \textbf{64}, 184104 (2001)

\bibitem{Wang:1998}
Y. G. Wang, W. Kleemann, W. L. Zhong, and L. Zhang. Impurity-induced phase transition in quantum paraelectrics Phys. Rev. B \textbf{57}, 13343 (1998).


\bibitem{Spinelli}
A. Spinelli,  M. A. Torija , C. Liu, C. Jan and C. Leighton.  Electronic transport in doped SrTiO$_{3}$: Conduction mechanisms and potential applications. \emph{Phys. Rev. B} \textbf{81}, 155110 (2010)

\bibitem{Mott1990}
N. F. Mott. Metal-insulator transitions, second edition. Taylor and Francis, London (1990).

\bibitem{Edwards}
P. P. Edwards and M. J. Sienko. Universality aspects of the metal–nonmetal transition in condensed media. Phys. Rev. B 17, 2575 (1978).

\bibitem{Behnia}
K. Behnia, On mobility of electrons in a shallow Fermi sea over a rough seafloor. J. Physics: Condens. Matt. \textbf{27}, 375501 (2015).

\bibitem{Allen2013}
S. J. Allen, B. Jalan, S. Lee, D. G. Ouellette, G. Khalsa, J. Jaroszynski, S. Stemmer, and A. H. MacDonald. Conduction-band edge and Shubnikov–de Haas effect in low-electron-density SrTiO$_{3}$. Phys. Rev. B \textbf{88}, 045114 (2013)

\bibitem{DeLima:2015}
B. S. de Lima \emph{et al.} Interplay between antiferrodistortive, ferroelectric, and superconducting instabilities in Sr$_{1-x}$Ca$_{x}$TiO$_{3−\delta}$. Phys. Rev. B \textbf{91}, 045108 (2015)

\bibitem{Supplement}
Information on sample details and experimental methods can be found in the supplement.

\bibitem{Hewson:1993}
A. C. Hewson. The Kondo Problem to Heavy Fermions (Cambridge Univ. Press, Cambridge, 1993).


\bibitem{Shi:2013}
Y. Shi \emph{et al.} A ferroelectric-like structural transition in a metal. Nature Materials \textbf{12}, 1024 (2013).

\bibitem{Benedek:2016} N. A. Benedek and T. Birol. ‘Ferroelectric’ metals reexamined: fundamental mechanisms and design considerations for new materials. J. Mater. Chem. C \textbf{4}, 4000 (2016).


\bibitem{Kolodiazhnyi:2010}
T. Kolodiazhnyi, M. Tachibana, H. Kawaji, J. Hwang and E. Takayama-Muromachi. Persistence of Ferroelectricity in BaTiO$_{3}$ through the Insulator-Metal Transition. Phys. Rev. Lett. \textbf{104}, 147602 (2010).

\bibitem{Wang:2012}
Y. Wang, X. Liu, J. D. Burton, S. S. Jaswal and E. Y. Tsymbal. Ferroelectric Instability Under Screened Coulomb Interactions. Phys. Rev. Lett. \textbf{109}, 247601 (2012).

\bibitem{vandermarel:2011}
D. van der Marel, J. L. M. van Mechelen and I. I. Mazin, Common Fermi-liquid origin of $T^{2}$ resistivity and superconductivity in n-type SrTiO$_{3}$. Phys. Rev. B \textbf{84}, 205111 (2011).

\bibitem{Lin:2015}
X. Lin, B. Fauqu\'e and K. Behnia. $T^{2}$ resistivity in in a small single-component Fermi surface. Science \textbf{349}, 945 (2015).

\bibitem{Matsushita:2005}
Y. Matsushita, H. Bluhm, T. H. Geballe, and I. R. Fisher. Evidence for Charge Kondo Effect in Superconducting Tl-Doped PbTe. Phys. Rev. Lett. \textbf{94}, 157002 (2005).

\bibitem{Lin:2014b}
X. Lin \emph{et al.} Multiple nodeless superconducting gaps in optimally doped  SrTi$_{1−x}$Nb$_{x}$O$_{3}$. Phys. Rev. B \textbf{90}, 140508(R) (2014)

\bibitem{Bauer:2012}
E. Bauer and M. Sigrist, Editors, Non-Centrosymmetric Superconductors, Introduction and Overview, Springer(2012).

\end{thebibliography}

\begin{thebibliography}{XX}

\bibitem{Lin:2013}
X. Lin, Z. Zhu, B. Fauqu\'e and K. Behnia. Fermi Surface of the Most Dilute Superconductor. Phys. Rev. X \textbf{3}, 021002 (2013).

\bibitem{Lin:2014}
X. Lin, G. Bridoux, A. Gourgout, G. Seyfarth, S. Kr\"{a}mer, M. Nardone, B. Fauqu\'e and K. Behnia. Critical Doping for the Onset of a Two-Band Superconducting Ground State in  SrTiO$_{3−\delta}$. Phys. Rev. Lett. \textbf{112}, 207002 (2014).

\bibitem{Niermann:2014}
D. Niermann, C. P. Grams, M. Schalenbach, P. Becker, L. Bohatý, J. Stein, M. Braden, and J. Hemberger.
Domain dynamics in the multiferroic phase of MnWO$_4$
Phys. Rev. B \textbf{89}, 134412 (2014).

\bibitem{DeLima:2015} B. S. de Lima \emph{et al.} Interplay between antiferrodistortive, ferroelectric, and superconducting instabilities in Sr$_{1-x}$Ca$_{x}$TiO$_{3-\delta}$. Phys. Rev. B \textbf{91}, 045108 (2015)

\bibitem{Balashova:1996}
E. V. Balashova , V. V. Lemanov , R. Kunze , G. Martin, and M. Weihnacht.
Ultrasonic study on the tetragonal and muller phase in SrTiO$_3$.
Ferroelectrics \textbf{183}, 75 (1996).

\bibitem{Aso:1976} J.G. Bednorz and K.A. M\"uller.
Sr$_{1-x}$Ca$_{x}$TiO$_3$: An XY Quantum Ferroelectric with Transition to Randomness.
Phys. Rev. Lett. \textbf{52}, 2289 (1984).

\bibitem{Guo:2012}
Y.J. Guo, Y.Y. Guo, L. Lin, Y.J. Gao, B.B. Jin, L. Kang, and J.-M. Liu.
Mean-field theory of ferroelectricity in Sr$_{1-x}$Ca$_{x}$TiO$_3$ ($0 \leq x \leq 0.4$).
Phys. Rev. B \textbf{86}, 014202 (2012).

\bibitem{Salje:2013}
E.K.H. Salje, O. Aktas, and M.A. Carpenter, V.V. Laguta, and J.F. Scott.
Domains within Domains and Walls within Walls: Evidence for Polar Domains in Cryogenic SrTiO$_3$.
Phys. Rev. Lett. \textbf{111}, 247603 (2013).

\bibitem{Allen:2013}
S.J. Allen, B. Jalan, S. Lee, D.G. Ouellette, G. Khalsa, J. Jaroszynski, S. Stemmer and A.H. MacDonald.
Conduction-band edge and Shubnikov-de Haas effect in low-electron density SrTiO$_3$.
Phys. Rev. B. \textbf{88}, 045114 (2013).


\end{thebibliography}
\end{document}